\documentclass[
 reprint,
 superscriptaddress,
 amsmath,amssymb,
 aps,
 prb,
floatfix,
]{revtex4-2}

\usepackage{graphicx}
\usepackage{dcolumn}
\usepackage{bm}
\usepackage{braket}
\usepackage{upgreek}
\usepackage[T1]{fontenc}
\usepackage{booktabs}

\usepackage{makecell}
\usepackage[hidelinks]{hyperref}
\usepackage[all]{hypcap} 

\begin{document}
\makeatletter

\newcommand\thefontsize[1]{{#1 The current font size is: \f@size pt\par}}

\makeatother
\preprint{APS/123-QED}

\title{
\texorpdfstring{Strained Donor-Bound Excitons in $^{28}$Si}
{Strained Donor Bound Excitons in Si}
}

\author{David A. Vogl}\email{david.vogl@wsi.tum.de}
 \affiliation{Walter Schottky Institute and School of Natural Sciences, Technical University of Munich, Am Coulombwall 4, 85748 Garching, Germany}
 \affiliation{Munich Center for Quantum Science and Technology, Schellingstraße 4,
 80799 Munich, Germany}
\author{Noah L. Braitsch}
 \affiliation{Walter Schottky Institute and School of Natural Sciences, Technical University of Munich, Am Coulombwall 4, 85748 Garching, Germany}
\affiliation{Munich Center for Quantum Science and Technology, Schellingstraße 4,
 80799 Munich, Germany}
\author{Başak Ç. Özcan}
 \affiliation{Walter Schottky Institute and School of Natural Sciences, Technical University of Munich, Am Coulombwall 4, 85748 Garching, Germany}
 \affiliation{Munich Center for Quantum Science and Technology, Schellingstraße 4,
 80799 Munich, Germany}
\author{Niklas S. Vart}
 \affiliation{Walter Schottky Institute and School of Natural Sciences, Technical University of Munich, Am Coulombwall 4, 85748 Garching, Germany}
 \affiliation{Munich Center for Quantum Science and Technology, Schellingstraße 4,
 80799 Munich, Germany}
\author{M. L. W. Thewalt}
 \affiliation{Department of Physics, Simon Fraser University, Burnaby, British Columbia, Canada V5A 1S6}
\author{Martin S. Brandt}
 \affiliation{Walter Schottky Institute and School of Natural Sciences, Technical University of Munich, Am Coulombwall 4, 85748 Garching, Germany}
 \affiliation{Munich Center for Quantum Science and Technology, Schellingstraße 4,
 80799 Munich, Germany}

\date{\today}

\begin{abstract}
We present a comprehensive experimental study of the neutral donor to donor-bound exciton transition (D$^0$$\rightarrow\,$D$^0$X) in isotopically enriched $^{28}$Si, focusing on the group-V donors P, As, and Sb under finely tuned uniaxial stress along the [100] and [110] crystal axes and magnetic fields from 3.5~mT to 1.7 T. From these measurements, donor-specific deformation potentials are extracted. The uniaxial electron deformation potential $\Xi_\mathrm{u}$ is found to be significantly larger than values reported for other states or transitions in silicon and shows a clear dependence on the donor species, indicating an increased sensitivity of the D$^0$X state to strain and central-cell effects. We also observe a magnetic field dependence of the hole shear deformation potential $d$, suggesting a more complex strain coupling mechanism than captured by standard theory. Diamagnetic shift parameters determined from Zeeman spectra show good agreement with earlier measurements. Our results provide a refined parameter set critical for the design of silicon quantum devices based on D$^0$X transitions.
\end{abstract}

\maketitle

\section{\label{sec:Introduction}Introduction}
Donors in silicon are investigated as a powerful implementation of qubits for quantum computation \cite{Kane_1998, https://doi.org/10.1002/qute.202000005}. They pose significant advantages over other leading qubit technologies, like superconducting transmon qubits \cite{PhysRevA.76.042319}, electro-statically defined quantum dots \cite{RevModPhys.85.961} or ion traps \cite{HAFFNER2008155}: The donor Coulomb potential offers a natural confinement of the electron, and atomically precise placement \cite{PhysRevLett.91.136104} in the silicon lattice promises inherent scalability. Owing to the advances in isotopic purification \cite{PhysRevLett.106.030801}, both the electronic and the nuclear spin of the donor show exceedingly long coherence times \cite{Tyryshkin_2011, Steger_2012, Saeedi_2013}. The high nuclear spin of the heavier group V donors has successfully been implemented as a higher dimensional qudit \cite{Asaad2020, FernandezDeFuentes2024, Yu2025}, and one and two qubit operations on individual donors have been demonstrated repeatedly \cite{Morello_2010, Pla_2013, Ma_dzik_2021, M_dzik_2022}. Fast initialization of the spin state or control of the donor charge state, however, remain challenging.\par
The optical excitation of neutral donors (D$^0$) to donor-bound excitons (D$^0$X) has been demonstrated as a reliable method for both, electron and nuclear spin state initialization \cite{PhysRevLett.97.227401, 10.1063/1.3577614} and hybrid opto-electrical readout \cite{Lo_2015, PhysRevApplied.11.054014} of ensembles of donors. It is based on spin-selective excitation and successive spin-to-charge conversion due to the dominant Auger recombination \cite{https://doi.org/10.1002/pssb.2220840216}. Recently, the D$^0$X readout of as few as 10$^4$ donors has been achieved \cite{conti_2024}, however, the application to individual donors has yet to be demonstrated. One major challenge in scaling down this readout technique to single donors is presented by its sensitivity to strain, leading to a strong spectral shift, broadening or splitting and, in extreme cases, dissociation of the donor-bound exciton complex, when the effect of local strain gradients exceeds the binding energy \cite{Lo_2015}.\par
Strain is a fundamental aspect of any nanodevice, as the mismatch in thermal expansion coefficients between different materials inherently introduces strain in the structure at cryogenic temperatures \cite{PhysRevApplied.9.044014, Asaad2020}. Spin qubits in Si/SiGe heterostructures even rely on the strained environment generated by the lattice mismatch for the confinement of the electronic states \cite{Fang_2023}. The sensitivity to strain of the D$^0$$\rightarrow\,$D$^0$X transition may also be used for in-situ strain mapping of silicon nanodevices, as recently demonstrated \cite{conti_2024}.\par
Hence, precise knowledge of the strain dependence of the D$^0$X excitation spectrum is desirable. In this context, the impact of strain resulting from implantation damage has been examined in detail \cite{https://doi.org/10.1002/qute.201800038}. Early investigations explored the effect of strain on the photoluminescence of bound multi-exciton complexes in the absence of an external magnetic field \cite{PhysRevLett.41.808, PhysRevB.45.11736}. More recently, the effects of strain on D$^0$X transitions in the presence of a magnetic field were studied, where measurements however were limited either to a single, fixed strain value \cite{Lo_2015, PhysRevMaterials.7.016202} or to very few data points and small overall stress \cite{conti_2024}. \par
This work presents more detailed experimental data for the P, As, and Sb neutral donor to donor-bound exciton transition (D$^0$$\rightarrow\,$D$^0$X) under [100] and [110] in-situ variable stress at various magnetic fields. Within each dataset, the established deformation potential theory describes the spectra well, yielding deformation potentials that are mostly consistent with earlier work. However, when comparing across magnetic fields, we find a clear dependence of the hole shear deformation potential $d$ on field strength, suggesting a previously neglected higher-order coupling between magnetic field, shear strain, and the D$^0$X hole states. In addition, we obtain a significantly larger electron deformation potential $\Xi_\mathrm{u}$ than reported by other methods, which may also be donor-dependent.\par
\section{\label{sec:Methods}Methods}
\subsection*{Electrical Detection of \texorpdfstring{D$^0$X}{D0X}}
Following optical excitation of the D$^0$$\rightarrow\,$D$^0$X transition, the donor-bound excitons decay predominantly via Auger recombination \cite{https://doi.org/10.1002/pssb.2220840216}, in which one of the electrons recombines with the hole, and transfers its energy to the second electron. This second electron is excited high into the conduction band, leaving behind an ionized donor. Although the excess conduction band electrons rapidly thermalize and are recaptured by the ionized donors, this process still temporarily increases the free charge carrier concentration, allowing for the electrical detection of optically excited donor-bound excitons through impedance measurements.

\subsection*{Samples}
Our samples are cut from a bulk $^{28}$Si crystal isotopically purified to 99.991\% as part of the Avogadro project \cite{PhysRevLett.106.030801}. They have dimensions of $2\times 2\times 8$ mm$^3$ and contain about $[^{31}$P$] = (1.5\pm0.2)\times 10^{14}$ cm$^{-3}$, $[^{75}$As$] = (3.4\pm1.1)\times 10^{14}$ cm$^{-3}$, and $[^{121}$Sb$] = (1.0\pm0.2)\times 10^{14}$ cm$^{-3}$. The boron content has not been measured, however the samples are known to be n-type.
\subsection*{Experimental Setup}
\begin{figure}[htbp]
    \centering
    \includegraphics{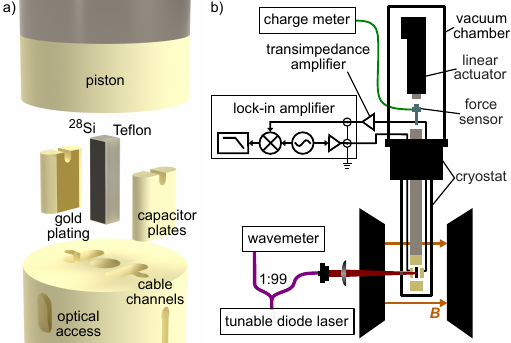}
    \caption{Schematic of the experimental setup for the detection of Auger electrons generated by the recombination of optically excited D$^0$X under mechanically applied stress. a) Exploded view rendering of the sample and its surrounding elements. All components colored in yellow are manufactured from PEEK. b) Depiction including laser, magnet, cryostat, lock-in amplifier, and actuator.}
    \label{fig:setup}
\end{figure}

Each sample is loosely placed in the sample holder custom made from polyether ether keton (PEEK) sketched in Fig.~\ref{fig:setup}a). It allows for optical access, and also holds in place capacitor plates used for the impedance measurements. This way we avoid inherent strain due to a tight mount or electrical contacts placed directly on the sample \cite{PhysRevApplied.11.054014}. The sample is further wrapped in thin pieces of Teflon tape to smooth out surface imperfections and thus distribute the applied stress homogeneously over the top and bottom surfaces.
The sample is cooled to just above liquid helium temperature in an Oxford Instruments CF935 helium flow cryostat and placed between the pole caps of a Bruker B-E25 electromagnet, as can be seen in the schematic of the experimental setup in Fig.~\ref{fig:setup}b). \par
For the excitation of the donor-bound excitons we use a Toptica continuously tunable diode laser with a coarse tuning range of 1010 - 1100 nm, a fine tuning range of 210 pm, and a linewidth specified to less than $10$ kHz. The laser wavelength is stabilized with the help of a High Finesse WS-7 wavemeter, which also allows us to measure the excitation wavelength with high precision.\par
We probe the impedance of the sample with an AC voltage at a test frequency of $f_\mathrm{test} = 903.14$ kHz generated by a Zurich Instruments UHF lock-in amplifier. The AC voltage is applied to one of the capacitor plates, and the electric signal on the other plate is then fed into a Femto DHPCA-100 transimpedance amplifier and finally back into the lock-in amplifier for demodulation. This way, we obtain a measure of the AC conductivity of the test structure containing the sample at the frequency $f_{\mathrm{test}}$.\par
Compressive stress is applied to the sample from above using an extended piston with a flat surface, driven by a Physics Instruments M-238.5PL high-load DC gear actuator. An integrated non-contact optical linear encoder measures the piston position to a precision of 100 nm. Simultaneously, the force present in the piston is measured with a Kistler 9217A piezo-electric force sensor together with an appropriate Kistler 5015A charge meter. In order to avoid friction introduced by a vacuum feedthrough, the actuator and piston are inside a vacuum chamber directly connected to the cryostat vacuum that is required for the helium.
\section{\label{sec:Theory}Theory}
The shift of the D$^0$$\rightarrow\,$D$^0$X transition energy $\Delta E$ under applied stress $\sigma$ and magnetic field $\bm{B}$ can be approximated by the combined changes in energy of the additional electron and hole \cite{Lo_2015, ross_phd, conti_2024}
\begin{equation}
    \Delta E_{\mathrm{D}^0\rightarrow\mathrm{D}^0\mathrm{X}}(\epsilon, \bm{B}) \approx \Delta E_\mathrm{e}(\epsilon, \bm{B}) + \Delta E_\mathrm{h}(\epsilon, \bm{B}),
\end{equation}
where $\epsilon$ denotes the strain associated with $\sigma$, and changes in the electron-electron and electron-hole interactions are neglected. Following effective mass theory,
we describe the weakly bound electron and hole based on the characteristics and symmetry of the conduction band (CB) minimum and valence band (VB) maximum, respectively. \par
\subsection*{Effects of Strain on Electron and Hole States}
The CB minimum in silicon is sixfold degenerate. In the presence of the donor potential, however, the translational symmetry of the crystal is broken, and the degeneracy is lifted by the valley-orbit (VO) coupling \cite{PhysRev.98.915}. In agreement with group theory, the states are then split into a non-degenerate ground state $A_1$ and two-fold and three-fold degenerate excited states $E$ and $T_2$, respectively, according to the irreducible representations of the appropriate point group $T_\mathrm{d}$. This VO coupling is commonly parameterized by an overall binding energy in the absence of the VO coupling $E_0$, and valley couplings $\Delta_1$ and $\Delta_2$ for neighboring and opposing valleys, respectively. Writing the VO coupling in the basis of the six individual CB valleys then yields the valley-orbit Hamiltonian
\begin{equation}
    \mathcal{H}_\mathrm{VO} = \begin{pmatrix}
    E_0 & \Delta_1 & \Delta_2 & \Delta_2 & \Delta_2 & \Delta_2 \\
    \Delta_1 & E_0 & \Delta_2 & \Delta_2 & \Delta_2 & \Delta_2 \\
    \Delta_2 & \Delta_2 & E_0 & \Delta_1 & \Delta_2 & \Delta_2 \\
    \Delta_2 & \Delta_2 & \Delta_1 & E_0 & \Delta_2 & \Delta_2 \\
    \Delta_2 & \Delta_2 & \Delta_2 & \Delta_2 & E_0 & \Delta_1 \\
    \Delta_2 & \Delta_2 & \Delta_2 & \Delta_2 & \Delta_1 & E_0 \\
    \end{pmatrix}\,.
\end{equation}\par
In the presence of strain, valley energies are shifted. Due to symmetry, a volume deformation potential $\Xi_\mathrm{d}$ and a uniaxial deformation potential $\Xi_\mathrm{u}$ are sufficient to describe this shift \cite{PhysRevB.45.11736}. The valley-strain interaction may then be solved together with the valley-orbit interaction to describe a repopulation of valleys (VR) due to strain as
\begin{equation}
    \mathcal{H}_\mathrm{VR} = \mathcal{H}_\mathrm{VO}\! + \Xi_\mathrm{d}\!\sum_i \epsilon_{ii} + \Xi_\mathrm{u}\!\left(\begin{smallmatrix}
         \epsilon_{11} &&&&&0 \\
        & \epsilon_{11} &&&& \\
        && \epsilon_{22} &&& \\
        &&& \epsilon_{22} && \\
        &&&& \epsilon_{33} & \\
        0&&&&& \epsilon_{33}
    \end{smallmatrix}\right),
    \vspace{0.1em}
\end{equation}
with $\epsilon_{ii}$ the diagonal components of the strain tensor. Strain shifts the energy of all involved states, and, depending on the symmetry of the applied stress, may lift the degeneracy of the excited $E$ and $T_2$ states to varying degrees.\par
Since the excited states are sufficiently separated in energy from the ground state, the latter is the only state relevant for the transitions measured in our experiment. However, any stress that is not oriented along the [111] crystal direction mixes the $E$ states into the $A_1$ ground state. Due to this mixing, the ground state solution to the valley repopulation Hamiltonian $\mathcal{H}_\mathrm{VR}$ under arbitrary strain yields lengthy analytical solutions. Instead, we present results of the ground state energy only for highly symmetric strain tensors corresponding to stress applied along the [100] and [110] crystal directions, as implemented in the experiment. The strain tensors for stress $\sigma$ along [100] and [110] are
\begin{gather}
    \epsilon^{100} = \begin{pmatrix}
        s_{11} & 0 & 0 \\
        0 & s_{12} & 0 \\
        0 & 0 & s_{12}
    \end{pmatrix}\sigma
    \\
        \epsilon^{110} = \begin{pmatrix}
        (s_{11}+s_{12})/2 & s_{44}/4 & 0 \\
        s_{44}/4 & (s_{11}+s_{12})/2 & 0 \\
        0 & 0 & s_{12}
    \end{pmatrix}\sigma\,,
\end{gather}
where $s_{11} = 7.68$ TPa$^{-1}$, $s_{12} = -2.14$ TPa$^{-1}$, and $s_{44} = 12.6$ TPa$^{-1}$ are the stiffness parameters of silicon \cite{mason1958physical} and $\sigma < 0$ for compressive stress.
The solutions for the shift of the electron energy with strain are then given by
\begin{widetext}
\begin{equation}
    \label{eq:electron_strain_shift}
\begin{alignedat}{2}
    \Delta E_\mathrm{e}(\epsilon^{100}) &= 
    (\Xi_\mathrm{d}+\frac{1}{3}\Xi_\mathrm{u})\epsilon_\mathrm{a}
    +2&&\Xi_\mathrm{u}\epsilon_\mathrm{b}
    -3\Delta_2
    -3 \sqrt{
        \Delta _2^2
        -4\Delta _2  \Xi _\mathrm{u} \epsilon_\mathrm{b}/3
        + 4\Xi _\mathrm{u}^2\epsilon_\mathrm{b}^2
        }
    \\
    \Delta E_\mathrm{e}(\epsilon^{110}) &= 
 (\Xi_\mathrm{d}+\frac{1}{3}\Xi_\mathrm{u})\epsilon_\mathrm{a}
  -\phantom{2}&&\Xi _\mathrm{u}\epsilon _\mathrm{b}
 -3 \Delta _2
 -3\sqrt{
    \Delta _2^2
    + 2 \Delta _2 \Xi _\mathrm{u}  \epsilon _\mathrm{b} /3
    +  \phantom{4}\Xi _\mathrm{u}^2 \epsilon _\mathrm{b}^2
    }\,,
\end{alignedat}
\end{equation}
\end{widetext}
where we defined
\begin{equation}
    \epsilon_\mathrm{a} = (s_{11}+2s_{12})\sigma \quad\mathrm{and}\quad \epsilon_\mathrm{b} = \frac{1}{12}(s_{11}-s_{12})\sigma.
\end{equation}
For small stress such that $\Xi_\mathrm{u}\epsilon_\mathrm{b} \ll \Delta_2$, to a good approximation, all terms containing $\epsilon_\mathrm{b}$ cancel, and we are left with a linear energy shift with strain $\epsilon_\mathrm{a}$.\par
\begin{figure}[htbp]
    \includegraphics{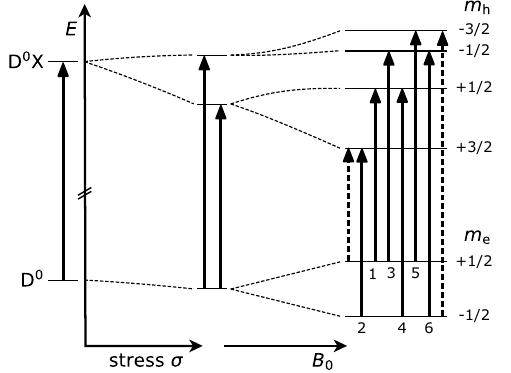}
    \caption{\label{fig:transitions} Schematic of energy levels involved in the D$^0$$\rightarrow\,$D$^0$X transition as a function of uniaxial compressive stress $\sigma$ and magnetic field strength $B_0$. Transitions are labeled following the usual convention \cite{Steger_2012}, which reflects their energy ordering in a magnetic field without strain. Applied strain, however, can modify this ordering, as shown. Note that even though $m_\mathrm{h}$ is not a good quantum number in the presence of both strain and magnetic field, we still use it as a label for the hole states according to the high-field limit.}
\end{figure}

The uppermost VBs in silicon consist of two-fold degenerate heavy hole and light hole bands, characterized by hole magnetic quantum numbers $m_\mathrm{h}=\pm3/2$ and $m_\mathrm{h}=\pm1/2$, respectively, and both corresponding to a total angular momentum of $J=3/2$. At the VB maximum, these two bands are degenerate, but strain lifts this degeneracy and splits the VB maximum states into the heavy-hole and light-hole doublets. An expression for the energy shift as a function of strain may again be found purely by symmetry arguments, yielding the Pikus-Bir Hamiltonian \cite{BIR19631467}
\begin{equation}
    \hat{\mathcal{H}}_\mathrm{PB} = 
    a\sum_i \epsilon_{ii} + 
    b\sum_i \left(\hat{J}_i^2 - \frac{5}{4}\right)\epsilon_{ii} + 
    \frac{d}{2\sqrt{3}}\sum_{i\neq j} \hat{J}_i\hat{J}_j\epsilon_{ij}\,,
\end{equation}
where $a$, $b$, and $d$ are deformation potentials and $\hat{J}_i$ are the angular momentum projection operators. Explicit solutions to this Hamiltonian for [100]- and [110]-stress are given by
\begin{align}
    \Delta E_\mathrm{h}(\epsilon^{100}) &= \left\{
    \begin{array}{ll}
    -a\epsilon_\mathrm{a} -12b\epsilon_\mathrm{b}& \quad m_h\pm3/2, \\
    -a\epsilon_\mathrm{a} +12b\epsilon_\mathrm{b} & \quad m_h\pm1/2
    \end{array}
    \right. \\
        \Delta E_\mathrm{h}(\epsilon^{110}) &= \left\{
    \begin{array}{ll}
    -a\epsilon_\mathrm{a} -\sqrt{36b^2\epsilon^2_\mathrm{b} + d^2\epsilon^2_\mathrm{c}}& \quad m_h\pm3/2, \\
    -a\epsilon_\mathrm{a} +\sqrt{36b^2\epsilon^2_\mathrm{b} + d^2\epsilon^2_\mathrm{c}} & \quad m_h\pm1/2
    \end{array}\,,
    \right.
\end{align}
where we introduced $\epsilon_\mathrm{c} = \frac{1}{4}s_{44}\sigma$. The shift and splitting of the D$^0$ and D$^0$X states with strain due to the changes in the electron and hole energy is shown in Fig.~\ref{fig:transitions}.
\subsection*{Effects of Magnetic Fields on Electron and Hole States}
When including an external magnetic field $\bm{B}$, the electronic Zeeman interaction has to be considered for both, the initial and final state of the D$^0$$\rightarrow\,$D$^0$X transition. The neutral donor state D$^0$ only has one s-like electron with total angular momentum $S=1/2$. The Zeeman interaction thus splits the ground state into two levels with electron magnetic quantum number $m_\mathrm{e}=\pm1/2$ according to the electron Zeeman Hamiltonian
\begin{equation}
    \hat{\mathcal{H}}_\mathrm{eZ} = g_\mathrm{e}\upmu_\mathrm{B}\bm{B}\!\cdot\!\bm{\hat{S}},
\end{equation}
where $g_\mathrm{e}$ is the donor-specific electron g-factor, $\upmu_\mathrm{B}$ is the Bohr magneton, and $\bm{\hat{S}}$ is the spin vector operator. Since the VR model does not couple to the spin degree of freedom, the electron Zeeman Hamiltonian may be solved independently to yield
\begin{equation}
\label{eq:electron_Zeeman_shift}
    \Delta E_\mathrm{eZ}(B_0) = \pm\frac{1}{2}g_\mathrm{e}\upmu_\mathrm{B}B_0,
\end{equation}
with $B_0$ the magnitude of the magnetic field $\bm{B}$.\par
In the final D$^0$X state, the two electrons occupy the same state \cite{KIRCZENOW1977713}, and thus their spin cancels, so that the Zeeman interaction is governed by the $J=3/2$ hole. Due to spin-orbit coupling to the $J=1/2$ band, the hole Zeeman Hamiltonian takes a non-trivial form \cite{PhysRev.102.1030, PhysRevB.6.3836}
\begin{equation}
        \hat{\mathcal{H}}_\mathrm{hZ} = 
        g_1\upmu_\mathrm{B}\bm{B}\!\cdot\!\bm{\hat{J}} + 
        g_2\upmu_\mathrm{B}\bm{B}\!\cdot\!\bm{\hat{J}}^3,
\end{equation}
with g-factors $g_1$ and $g_2$ and the angular momentum vector operator $\bm{\hat{J}}$ for $J=3/2$. In itself, this spin-orbit coupling yields different g-factors for heavy holes and light holes
\begin{equation}
    \Delta E_\mathrm{hZ}(B_0) = \left\{
    \renewcommand{\arraystretch}{2.0}
    \begin{array}{ll}
    \pm\frac{3}{2}g_\mathrm{hh}\upmu_\mathrm{B}B_0 & \quad m_\mathrm{h}\pm3/2,\\
    \pm\frac{1}{2}g_\mathrm{lh}\upmu_\mathrm{B}B_0 & \quad m_\mathrm{h}\pm1/2,
    \end{array}
    \right.
\end{equation}
where $g_\mathrm{hh}$ and $g_\mathrm{lh}$ are functions of $g_1$, $g_2$, and the direction of the magnetic field relative to the crystal axes.\par
The Zeeman interaction lifts all remaining degeneracies, resulting in two Zeeman-split neutral donor ground states, and four split donor-bound exciton states, for a total of eight possible transitions (see Fig.~\ref{fig:transitions}). Only six of these transitions are dipole allowed to first order according to the selection rule $\Delta m = 0,\pm 1$. In the presence of both strain and magnetic field however, the hole angular momentum states are generally mixed \footnote{unless both stress and magnetic field are aligned with the same [100] crystal axis, which is not the case in our work}, and thus $m_\mathrm{h}$ is no longer a good quantum number. In this case, the selection rule may be partially relaxed, depending on the degree of mixing. For the same reason, an exact solution can only be found by solving the strain interaction simultaneously with the hole Zeeman interaction,
\begin{equation}\label{eq:total_hole_shift}
    \hat{\mathcal{H}}_\mathrm{PB}(\epsilon) + \hat{\mathcal{H}}_\mathrm{hZ}(\bm{B}) \quad\rightarrow\quad \Delta E_\mathrm{h}(\epsilon, \bm{B}),
\end{equation}
where solutions are not necessarily linear in strain or magnetic field.
\par
The external magnetic field also induces a diamagnetic shift of both the neutral donor and donor-bound exciton, which varies quadratically with $B_0$ \cite{Litvinenko_2016}. Within the resolution of our measurements, we were not able to resolve any difference in the diamagnetic shift of the hole states, and thus we model the total diamagnetic shift of the D$^0$$\rightarrow\,$D$^0$X transitions as
\begin{equation}
\label{eq:dia_shift}
    \Delta E_\mathrm{dia}(B_0) = \gamma B_0^2\,,
\end{equation}
with the diamagnetic shift parameter $\gamma$ independent of the specific initial or final state involved in the transition \cite{PhysRevB.7.4547}.\par
To summarize, our model for the total energy shift of the D$^0$$\rightarrow\,$D$^0$X transition constitutes the sum of all contributions as defined in Eqs.~\ref{eq:electron_strain_shift}, \ref{eq:electron_Zeeman_shift}, \ref{eq:total_hole_shift}, and \ref{eq:dia_shift}
\begin{multline}
\Delta E_{\mathrm{D}^0\rightarrow\mathrm{D}^0\mathrm{X}}(\epsilon, \bm{B}) = \Delta E_\mathrm{e}(\epsilon) - \Delta E_\mathrm{eZ}(B_0)\\  + \Delta E_\mathrm{h}(\epsilon, \bm{B}) +  \Delta E_\mathrm{dia}(B_0)\,,
\end{multline}
where the electron Zeeman term $\Delta E_\mathrm{eZ}$ enters with a negative sign, as it applies to the initial state of the transition.
\section{\label{sec:Results}Experimental Results}
\begin{figure}[htbp]
    \includegraphics{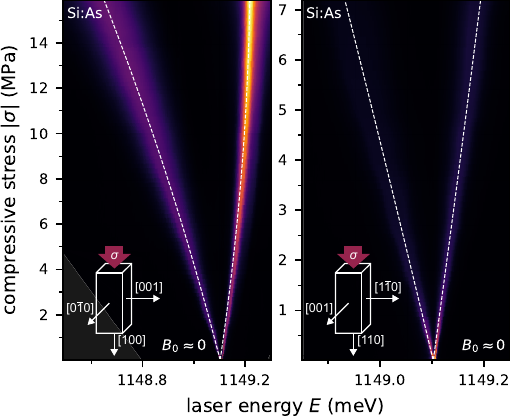}
    \caption{\label{fig:As_35G}Arsenic neutral donor to donor-bound exciton transition spectra (D$^0$$\rightarrow\,$D$^0$X) in isotopically enriched $^{28}$Si, observed by measuring the sample AC conductivity under varying uniaxial stress. The two panels correspond to stress applied along the [100] and [110] crystal axes, as labeled. Data is shown as 2D color plots using nearest-neighbor coloring without interpolation, and brighter colors indicate a stronger signal intensity. Grey regions indicate areas outside the acquired data range. The white dashed lines represent a global fit of deformation theory to an extended dataset, see text for details.}
\end{figure}

Figure \ref{fig:As_35G} presents color plots of the arsenic D$^0$$\rightarrow\,$D$^0$X transition energy $E_{\mathrm{D}^0\rightarrow\mathrm{D}^0\mathrm{X}}$ measured under variable uniaxial stress $\sigma$ along the [100] and [110] crystal directions. The nominal magnetic field $B_0$ was set as low as possible (limited to 3.5 mT by the remanence of our electromagnet). Insets in the bottom-left corner of each plot illustrate the orientation of applied stress or an external magnetic field relative to the crystal axes.\par 
A clear spectral broadening of the arsenic D$^0$$\rightarrow\,$D$^0$X transitions is evident, which we attribute to inhomogeneous stress across the sample volume. We find that the linewidth approximately scales as $\sigma \cdot{\partial E_{\mathrm{D}^0\rightarrow\mathrm{D}^0\mathrm{X}}}/{\partial \sigma}$. Due to this broadening, the signal intensity generally decreases at higher stress. However, this effect is reduced in regions where the transition energy becomes insensitive to stress in first order, as evidenced occasionally in our spectra when a line passes through an infinite slope.\par
For both stress orientations we observe a linear increase of the "zero-field" splitting with stress, as the light-hole and heavy-hole states become increasingly separated. 
The non-linear regime is reached only for the largest stresses in our measurements. Because Eq.~\ref{eq:electron_strain_shift} predicts a more pronounced non-linear effect for stress along [100],  we specifically extended our measurement range to higher stresses for the [100] orientation as shown here. Only for these measurements, we gradually increased the laser power with applied stress to compensate for the signal loss due to broadening.\par
\begin{figure*}[htbp]
    \includegraphics{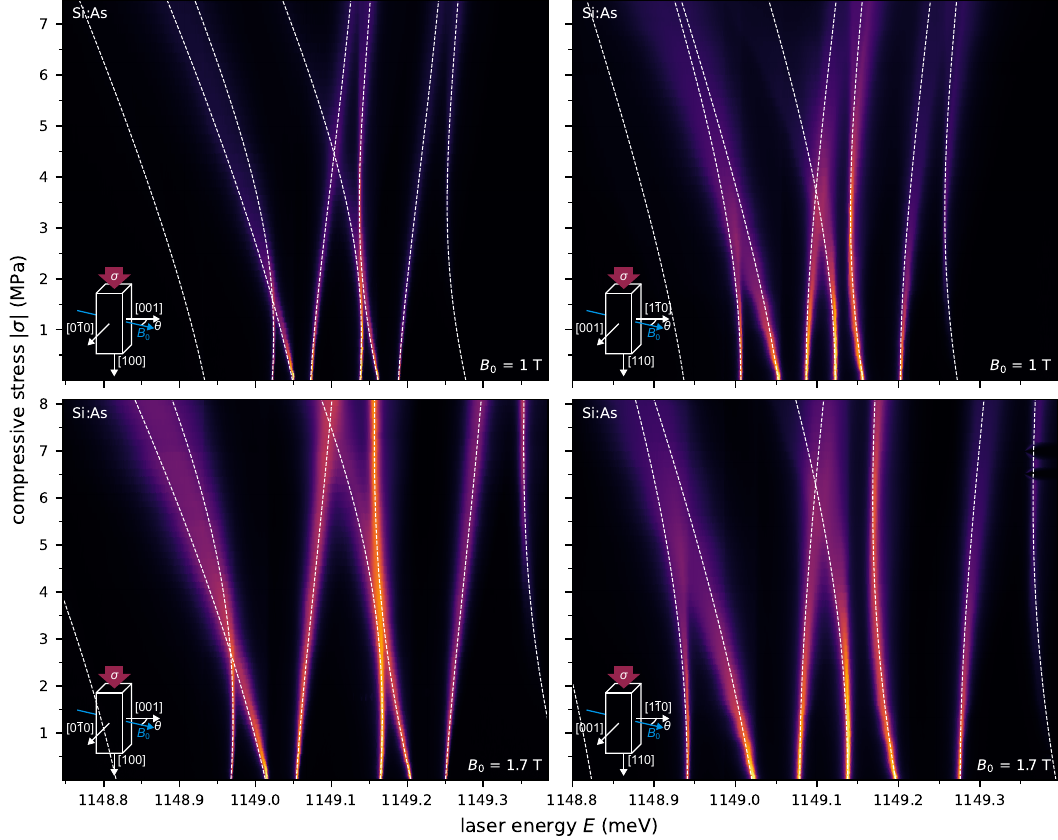}
    \caption{\label{fig:As_high_field_strain_maps}Arsenic neutral donor to donor-bound exciton transition spectra (D$^0$$\rightarrow\,$D$^0$X) in isotopically enriched $^{28}$Si under varying uniaxial stress and external magnetic fields. Each panel corresponds to a unique combination of stress orientation ([100] or [110]) and magnetic field strength (1.0 T or 1.7 T), as labeled in the subplots. For details regarding color code and fitted lines, refer to the caption of Fig.~\ref{fig:As_35G}.}
\end{figure*}
Figure \ref{fig:As_high_field_strain_maps} shows similar measurements for both stress orientations, but now with the addition of constant magnetic fields of 1.0 T and 1.7 T. The applied field lifts the remaining degeneracies, revealing up to eight distinct transitions. The eight transitions can be grouped within the four-dimensional hole state subspace, which is duplicated and shifted by the Zeeman splitting of the two possible electron spin states $m_\mathrm{e}=\pm1/2$. Thus, for any transition, a corresponding transition can be identified, that is only shifted by a constant amount equal to the electron Zeeman splitting. Any crossing or avoided crossing between transitions that belong to different electron spin states $m_\mathrm{e}$ is therefore accidental, and has no physical significance.\par
While the two outermost transitions are nominally forbidden to first order, this condition is relaxed by the introduction of strain, since the hole eigenstates are no longer pure eigenstates of angular momentum. Indeed, we detect a significantly enhanced intensity for the higher-energy forbidden line under strain.\par
A total of 13 data sets for phosphorus and 13 data sets for arsenic were analyzed independently. Each set consists of spectral maps for both stress orientations and magnetic fields of 3.5 mT, 0.35 T, 0.5 T, 1.0 T, and 1.7 T, including repeated measurements for [100] stress at 3.5 mT, where the maximum applied stress was gradually increased. For clarity, the measurements with lower maximum stress are not shown. The arsenic data for 0.35 T and 0.5 T, along with all phosphorus data, are presented in the Appendix in Figs.~\ref{fig:As_low_field_strain_maps} and \ref{fig:Ph_all_strain_maps}.\par

The P and As data sets were each fitted with a single global least-squares fit to our model defined in Sec.~\ref{sec:Theory}. The resulting fits are shown in all spectral maps as white dashed lines. For the fits, the electron and hole g-factors and valley-orbit couplings were assumed fixed as given in Tab.~\ref{tab:fixed_params}, and the primary free parameters were given by the deformation potentials for the electrons and holes, which were generally shared between all data sets.
\begin{table}
\caption{\label{tab:fixed_params}Electron and hole g-factors and valley-orbit couplings as used in the global fit of P and As data and the modeling of Sb data.}
\renewcommand{\arraystretch}{1.1}
\begin{ruledtabular}
\begin{tabular}{cccc}
& $g_\mathrm{e}^{\,\mathrm{a}}$ & $(g_1, g_2)^{\,\mathrm{b}}$ & ${\Delta_2}^{\,\mathrm{c}}$ (mV) \\
\midrule
P & 1.99850 & (0.82, 0.22) & -2.17 \\
As & 1.99837 & (0.82, 0.22) & -3.75 \\
Sb & 1.99858 & (0.82, 0.22) & -2.05
\end{tabular}
\end{ruledtabular}
\begin{minipage}{\linewidth}
\vspace{0.5ex}
\flushleft
\footnotesize{$^\mathrm{a}\!$ \cite{PhysRev.114.1219}, $^\mathrm{b}\!$ \cite{franke_phd, Lo_2015, kaminskii1980}, $^\mathrm{c}\!$ \cite{PhysRevB.23.2082}.}
\end{minipage}
\vspace{-5ex}
\end{table}
The only exception was the hole shear deformation potential $d$, for which we found it necessary to allow a magnetic field dependence in order to appropriately fit our complete data set. The remaining free parameters were an independent energy offset $E_0$ and the angle $\theta$ between the magnetic field and the crystal [001] and [1$\bar{1}$0] axes for stress oriented along [100] and [110], respectively. We found that our data, and also the Zeeman data presented later, are well explained by the hole g-factors taken from literature for angles $\theta_{100} = (16.2\pm0.6)^{\circ}$ and $\theta_{110} = (5.8\pm1.2)^{\circ}$. These angles are reasonable considering that the samples were aligned only by eye, and small rotations of the sample holder are expected when inserting the holder into the cryostat.\par
\begin{table*}
\caption{\label{tab:ex}Deformation potentials in eV as determined here and published previously.}
\renewcommand{\arraystretch}{1}
\begin{ruledtabular}
\begin{tabular}{cccccc}
 & \multicolumn{2}{c}{Present work (D$^0$$\rightarrow\,$D$^0$X)} & \multicolumn{3}{c}{Previous results} \\\\
 & P & As & P$^0$$\rightarrow\,$P$^0$X & \multicolumn{2}{c}{Other states/transitions} \\
  & exp. & exp. & exp. & exp. & theor. \\
\midrule
$\Xi_\mathrm{u}$ & 
    $17.41\pm0.05$ & 
    $19.05\pm0.07$ & 
    15.5\footnotemark[1] & 
    \makecell[c]{$8.5\pm0.2$\footnotemark[5]\\$8.6\pm0.2$\footnotemark[7]\\7.7\footnotemark[8]\\$8.3\pm0.3$\footnotemark[9]\\$11\pm1$\footnotemark[11]\\$8.77\pm0.07$\footnotemark[12]\\$8.6\pm0.4$\footnotemark[13]\\$8.5\pm0.1$\footnotemark[14]\\$11.4\pm1.1$\footnotemark[15]\\$8.6\pm0.4$\footnotemark[16]} & 
    \makecell[c]{10.5\footnotemark[17]\\9.29\footnotemark[18]\\8.86\footnotemark[19]\\9.16\footnotemark[20]\\8.47\footnotemark[22]\\8.0\footnotemark[23]}\\
    \\
$\Xi_\mathrm{d} + \frac{\Xi_\mathrm{u}}{3} - a$ & 
    $1.39\pm0.03$ & 
    $1.43\pm0.04$ & 
    2.0\footnotemark[1] & 
    \makecell[c]{$3.8\pm0.5$\footnotemark[7]\\$1.5\pm0.3$\footnotemark[13]} & 
    \makecell[c]{2.5\footnotemark[17]\\0.29\footnotemark[18]\\-0.77\footnotemark[19]\\1.72\footnotemark[20]\\1.79\footnotemark[22]\\1.2\footnotemark[23]}\\
    \\
$b$ & 
    $-1.7719\pm0.0019$ & 
    $-1.8179\pm0.0016$ & 
    \makecell[c]{-1.7\footnotemark[1]\\$-1.5\pm0.2$\footnotemark[2]\\-7\footnotemark[3]\\$-1.72\pm0.05$\footnotemark[4]} & 
    \makecell[c]{$-1.49\pm0.05$\footnotemark[5]\\$-1.35\pm0.05$\footnotemark[6]\\$-2.4\pm0.2$\footnotemark[7]\\$-1.36\pm0.13$\footnotemark[10]\\$-2.10\pm0.10$\footnotemark[13]} & 
    \makecell[c]{-2.33\footnotemark[17]$^,$\footnotemark[18]\\-2.21\footnotemark[19]\\-2.35\footnotemark[20]\\-2.58\footnotemark[21]\\-2.27\footnotemark[22]\\-2.18\footnotemark[23]}\\
    \\
$d$ & 
    \multicolumn{2}{c}{\makecell[c]{\begin{tabular}{rl}
    $B_0=0:\ -4.88\pm0.19$\\
    $B_0\rightarrow\infty:\ -3.78\pm0.12$
    \end{tabular}}} & 
    \makecell[c]{-5.1\footnotemark[1]\\$-4.2\pm0.3$\footnotemark[2]\\-4\footnotemark[3]\\$-4.53\pm0.1$\footnotemark[4]} & 
    \makecell[c]{$-4.08\pm0.10$\footnotemark[5]\\$-3.95\pm0.10$\footnotemark[6]\\$-5.3\pm0.4$\footnotemark[7]\\$-3.1\pm0.3$\footnotemark[10]\\$-4.85\pm0.15$\footnotemark[13]\\$-5.2\pm0.3$\footnotemark[14]} & 
    \makecell[c]{-4.75\footnotemark[17]\\-5.32\footnotemark[20]\\-5.40\footnotemark[21]\\-3.69\footnotemark[22]}\\
\end{tabular}
\end{ruledtabular}
\begin{minipage}[t]{0.50\textwidth}
\footnotetext[1]{Supplementary information of \cite{Lo_2015}, analyzed data was first presented in \cite{PhysRevLett.41.808}, measured at $B_0=0$.}
\footnotetext[2]{Ref.~\cite{conti_2024}, measured at $B_0=345$ mT.}
\footnotetext[3]{Fit to a single strain value, $B_0\in[0,\ 300]$ mT \cite{PhysRevMaterials.7.016202}.}
\footnotetext[4]{Photoluminescence spectra, $B_0=0$ \cite{PhysRevB.45.11736}.}
\footnotetext[5]{Photoluminescence spectra, boron-bound exciton, $B_0=0$ \cite{PhysRevB.45.11736}.}
\footnotetext[6]{Photoluminescence spectra, neutral boron acceptor, $B_0=0$ \cite{PhysRevB.45.11736}.}
\footnotetext[7]{Indirect optical absorption edge, $B_0=0$  \cite{PhysRev.143.636}.}
\footnotetext[8]{Temperature-dependent piezoresistance \cite{PhysRev.105.525}.}
\footnotetext[9]{Temperature-dependent piezoresistance \cite{PhysRev.130.1667}.}
\footnotetext[10]{Cyclotron resonance \cite{PhysRev.129.1041}.}
\footnotetext[11]{Group V donor hyperfine interaction \cite{PhysRev.124.1068}.}
\end{minipage}
\hfill
\begin{minipage}[t]{0.42\textwidth}
\footnotetext[12]{Donor piezospectroscopy \cite{PhysRevB.6.2348}.}
\footnotetext[13]{Indirect TO phonon-assisted exciton spectra, $B_0=0$ \cite{PhysRevB.3.2623}.}
\footnotetext[14]{Cyclotron resonance line width \cite{ITO196426}.}
\footnotetext[15]{Lithium donor hyperfine interaction \cite{PhysRevB.1.4071}.}
\footnotetext[16]{Effect of carrier concentration on elastic constants \cite{PhysRev.161.756}.}
\footnotetext[17]{Nonlocal empirical pseudopotential band structure \cite{10.1063/1.363052}.}
\footnotetext[18]{Nonlocal empirical pseudopotential band structure \cite{PhysRevB.48.14276}.}
\footnotetext[19]{Tight binding band structure \cite{PhysRevB.47.7104}.}
\footnotetext[20]{Density functional band structure \cite{PhysRevB.34.5621}.}
\footnotetext[21]{$k\!\cdot\! p$ method band structure \cite{PhysRev.142.530}.}
\footnotetext[22]{Local empirical pseudopotential band structure \cite{PhysRevB.39.7974}.}
\footnotetext[23]{LMTO band structure \cite{SCHMID199039}.}
\end{minipage}
\end{table*}
\section{\label{sec:Discussion}Discussion}
\subsection*{\label{sec:Deformation Potentials}Deformation Potentials}
\begin{figure}[htbp]
    \includegraphics{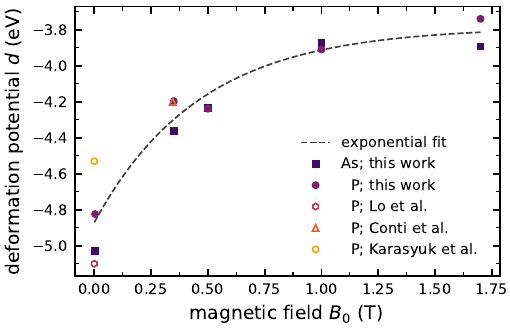}
    \caption{\label{fig:magnetic_d} Variation of the hole deformation potential $d$ with the applied external magnetic field strength $B_0$. Our presented deformation potentials are the result of a global fit to a large data set of the D$^0$$\rightarrow\,$D$^0$X transition under various uniaxial stress and magnetic fields. We also include previous results measured by Lo \textit{et al.}~\cite{Lo_2015}, Conti \textit{et al.}~\cite{conti_2024}, and Karasyuk \textit{et al.}~\cite{PhysRevB.45.11736}.}
\end{figure}

We find excellent agreement between our data and model. The best fit results for the deformation potentials are presented in Tab.~\ref{tab:ex} and Fig.~\ref{fig:magnetic_d}. Table \ref{tab:ex} also cites a number of previous results for the deformation potentials. It is important to note that while any lattice site point defect in the silicon lattice can be described by equivalent deformation potentials, this does not necessarily mean that they have to be equal. On the contrary, Pikus and Bir have shown that deformation potentials depend strongly on the wave function, and may vary by several tens of percent \cite{bir1974symmetry}. For this reason, the best comparison can only be made with other results for the same D$^0$$\rightarrow\,$D$^0$X transitions, for which however the available data is scarce and limited to phosphorus \cite{PhysRevB.45.11736, Lo_2015, conti_2024, PhysRevMaterials.7.016202}. Results from other methods or transitions may still provide some insight, as for weakly bound states, deformation potentials are expected to be similar. In particular, the deformation potentials for neutral donor states have been established in literature to be very similar to that of free electrons or excitons, as reported in Tab.~\ref{tab:ex}.\par
We find that our extracted hole deformation potentials $b$ and $d$ agree very well with previous results for the D$^0$$\rightarrow\,$D$^0$X transition, and to a lesser degree also with those obtained for other states/transitions. Within the experimental uncertainty, we observe no significant variation between the donor species. The hole deformation potential $a$ cannot be determined independently, as it has the same symmetry as certain contributions from the electron deformation potential. Instead, we can collect all hydrostatic terms as $\Xi_\mathrm{d}+\Xi_\mathrm{u}/3-a$, which again compares reasonably well with literature.\par
In contrast, the electron deformation potentials, and in particular the uniaxial component $\Xi_\mathrm{u}$, do not agree with literature values. Our measured $\Xi_\mathrm{u}$ is significantly larger than reported for other methods, but matches the only other result for the D$^0$$\rightarrow\,$D$^0$X transition reported by Lo \textit{et al.}~\cite{Lo_2015} for phosphorus, who also found a larger $\Xi_\mathrm{u}$. This leads us to believe that the larger $\Xi_\mathrm{u}$ is due to the specific nature of the D$^0$$\rightarrow\,$D$^0$X transition, and not merely a measurement or analysis error. Preliminary density functional theory (DFT) calculations indicate a stronger localization of the donor-bound exciton state compared to the neutral donor state \footnote{U. Gerstmann, private communication (2025).}, which may account for the observed increase in $\Xi_\mathrm{u}$. We further find that $\Xi_\mathrm{u}$ clearly differs for the two donor species, consistent with an enhanced sensitivity to central-cell effects in a more localized D$^0$X state, where chemical differences at the donor site become more pronounced.\par
Our results for a magnetic-field-dependent hole deformation parameter $d$ are summarized in Fig.~\ref{fig:magnetic_d}.
The deformation potential appears to first decrease strongly with the introduction of an external magnetic field, and then saturate at larger fields. Still, $d$ overall reduces to about 3/4 of its original value. A strong argument for a magnetic field dependence of $d$ for the D$^0$$\rightarrow\,$D$^0$X transition can be found when including previous results \cite{Lo_2015, conti_2024, PhysRevB.45.11736}. When considering the magnetic field at which those results were obtained, they align closely with the overall trend of our data, as is apparent from Fig.~\ref{fig:magnetic_d}. A simple exponential fit of our data for both donor species and the three literature values yields deformation potentials of
\begin{align*}
    d(B_0=0) &= (-4.88\pm0.19)\ \mathrm{eV} \\
    d(B_0\rightarrow\infty) &= (-3.78\pm0.12)\ \mathrm{eV},
\end{align*}
with a characteristic exponent of $\tau = (-2.2\pm0.7)$ 1/T. We emphasize that $d$ was introduced as a material specific constant in the Pikus-Bir Hamiltonian \cite{bir1974symmetry}, which by definition includes only contributions linear in strain and excludes magnetic field components. Consequently, the magnetic field dependence observed here has to be due to some interaction not included in our model. Still, from the fact that the magnetic field dependence of our results as well as those of Lo \textit{et al.}~\cite{Lo_2015}, Conti \textit{et al.}~\cite{conti_2024}, and Karasyuk \textit{et al.}~\cite{PhysRevB.45.11736} may be modeled by a variation of the hole shear deformation potential $d$, we can conclude that this interaction most likely involves a magnetic-field-dependent coupling of shear strain to the hole state of the D$^0$X. Such strain-magnetic~field interactions are indeed not generally forbidden by symmetry \cite{honerlage2018symmetry}. The saturation of the effect at larger magnetic fields may indicate that the interaction involves a mixing of hole states, which could become effectively reduced or decoupled at stronger fields.\par
\begin{figure*}[htbp]
    \includegraphics{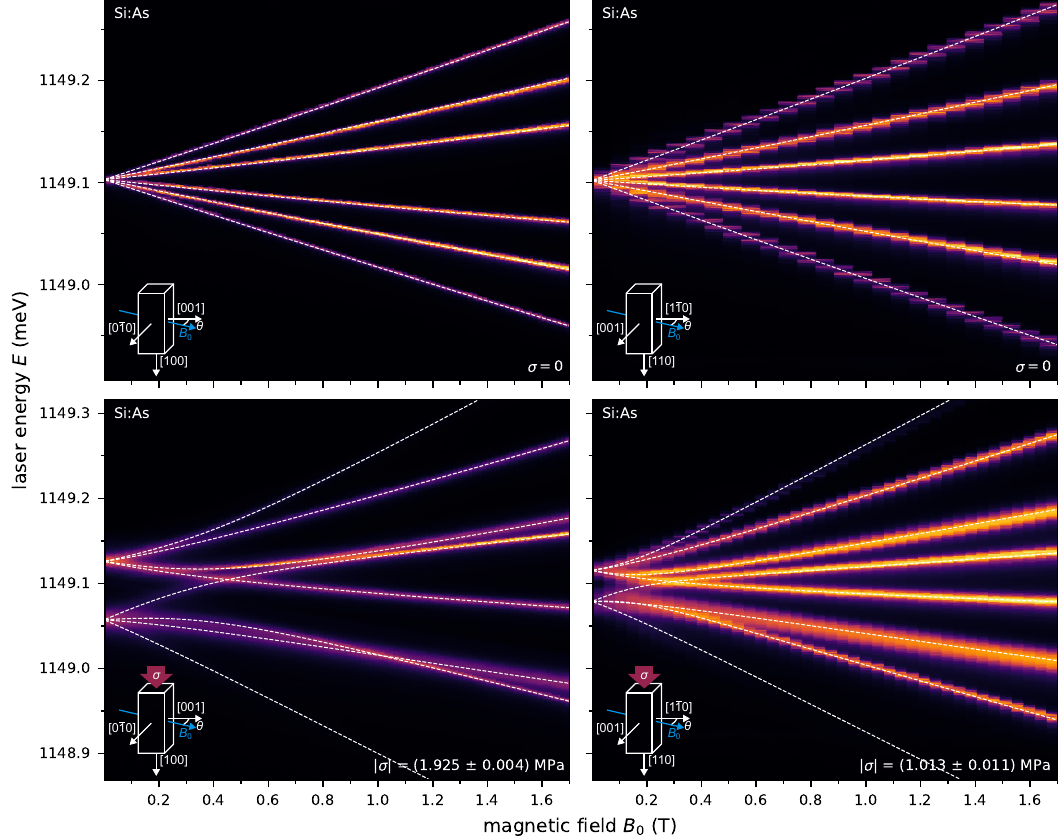}
    \caption{\label{fig:As_all_Zeemans}Zeeman maps of the arsenic neutral donor to donor-bound exciton transition (D$^0$$\rightarrow\,$D$^0$X) in isotopically enriched $^{28}$Si, observed by measuring the sample AC conductivity under varying magnetic fields. The four panels correspond to combinations of nominally zero or a small compressive stress, oriented along either [100] or [110] crystal axes, as labeled in the subplots. For details regarding color code and fitted lines, refer to the caption of Fig.~\ref{fig:As_35G}.}
\end{figure*}
\subsection*{Zeeman spectra}
Figure \ref{fig:As_all_Zeemans} shows Zeeman maps of the As D$^0$$\rightarrow\,$D$^0$X transition for the two sample orientations, under nominally zero strain and under a small, intentionally introduced strain, as typically encountered for D$^0$X experiments in slightly strained nanostructures \cite{mansir_phd, Asaad2020}. In the presence of this strain, a small zero-field splitting appears and the transition lines are broadened to various degrees, but individual lines can still be distinguished at relatively low magnetic fields. As previously observed, the higher-energy, nominally forbidden transition exhibits an enhanced intensity in a strained sample. However, this enhancement diminishes at larger magnetic fields, which decouple the angular momentum states, and thus restore the selection rules. Again, pairs of lines corresponding to the electron spin projections $m_\mathrm{e}=\pm1/2$ may be identified, all split by the same Zeeman splitting that, however, now increases linearly with the magnetic field strength $B_0$ and vanishes at zero magnetic field. The corresponding Zeeman maps for phosphorus are shown in the Appendix in Fig.~\ref{fig:Ph_all_Zeemans}.\par
We performed a second global fit of our model to all available Zeeman data, treating the diamagnetic shift parameter $\gamma$ as a shared fit parameter. For the intentionally strained data, the deformation potentials were fixed to our earlier results, which allowed us to accurately determine the applied stress in the Zeeman data while also maintaining excellent agreement between our fit and experimental data. Again, the only additional free parameters were the angle $\theta$, parameterizing a rotation of the sample in the sample holder, and a constant energy offset $E_0$. We extract diamagnetic shifts of $\gamma^\mathrm{As}=(1.96\pm0.04)$ $\upmu$eV/T$^2$ and $\gamma^\mathrm{P}=(2.24\pm0.04)$ $\upmu$eV/T$^2$, in good agreement with literature values \cite{PhysRevB.7.4547, Litvinenko_2016}.
\subsection*{Simulation of Sb data}
Our sample also contains a substantial amount of antimony, however the Sb D$^0$$\rightarrow\,$D$^0$X transition exhibits a much weaker signal in our measurements. This may be due to a much smaller oscillatory strength of the transition \cite{PhysRev.163.721}, or some other effect related to an equilibrium of donor and boron charge states that favors the ionized state of the Sb donor. For this reason, a notable Sb signal was only achieved for a much larger laser power, where the neighboring P transitions exhibit second order effects in the electrical detection. These appear as a narrower, inverted line shape superimposed on the primary signal, leading to an apparent splitting, as evident in Figs.~\ref{fig:Sb_all_strain_maps} and \ref{fig:Sb_all_Zeemans} shown in the Appendix.\par
The spectral proximity of the Sb transition to the corresponding P transition, a relatively weak signal and a generally strong broadening made a reliable fit of the Sb data challenging. Instead, we simulate the Sb D$^0$$\rightarrow\,$D$^0$X transitions using our model, with the P deformation potentials from Tab.~\ref{tab:ex}, the P diamagnetic shift parameter $\gamma^\mathrm{P}$ determined above, and the material specific parameters of Sb from Tab.~\ref{tab:fixed_params}. The simulation is shown as white dashed lines in Figs.~\ref{fig:Sb_all_strain_maps} and \ref{fig:Sb_all_Zeemans}, tracing the weaker Sb transitions next to the dominant P transitions. We find good agreement between the Sb data and this simulation, whereas simulations based on the As deformation potentials clearly did not reproduce the Sb spectra. We note that this is consistent with the donor binding energy and valley-orbit coupling being very similar for P and Sb, but different for As. \par
\section{\label{sec:Summary}Summary and Conclusions}
We have conducted a detailed study of the P, As, and Sb D$^0$$\rightarrow\,$D$^0$X transitions in isotopically enriched $^{28}$Si under variable uniaxial stress and external magnetic fields. We performed a global least-squares fit to a large dataset containing multiple magnetic field strengths from 3.5 mT to 1.7 T and stress oriented along the [100] and [110] crystal axes, and extracted deformation potentials for the D$^0$$\rightarrow\,$D$^0$X transition of P and As with high precision.\par
Our results show that while the hole deformation potentials are consistent with previous literature, the uniaxial electron deformation potential $\Xi_\mathrm{u}$ is donor-dependent and significantly larger than values reported for other experimental methods and theory. This suggests an enhanced sensitivity of the D$^0$X state to strain and central-cell effects, likely arising from increased spatial localization. Additionally, a magnetic field dependence of the hole deformation potential $d$, consistent with earlier literature on the D$^0$$\rightarrow\,$D$^0$X transition when accounting for field strength, suggests a nontrivial coupling of hole states to shear strain that is not explicitly captured in the basic deformation potential theory. Furthermore, the analysis of P and As Zeeman spectra enabled the determination of diamagnetic shift parameters, which were found to agree well with earlier results.\par
The relatively low intensity of the Sb D$^0$$\rightarrow\,$D$^0$X transitions, combined with their proximity to P, made a global fit with multiple parameters challenging. Instead we simulated the Sb spectra relying on the deformation potentials and diamagnetic shift parameters obtained for P, and found good agreement with the experimental Sb data. In contrast, deformation potentials obtained for As did not reproduce the Sb data.\par
Potential next steps include a more detailed theoretical study with the help of density functional theory or other ab initio methods to clarify the origin of the observed differences in deformation potentials, particularly for $\Xi_\mathrm{u}$. Extending the hole Hamiltonian to include higher-order strain terms could help account for magnetic-field-dependent variations in the hole deformation potential $d$. A more dedicated investigation of samples containing Sb and Bi would allow for an even more complete donor-specific parameter set, supporting both theoretical models and practical applications. Finally, shaping the sample ends in pyramidal form could further improve the strain homogeneity \cite{PhysRevB.45.11736}, reducing spectral broadening and allowing for even more precise measurements.\par
Our results point to a previously neglected strain-magnetic field interaction, that may be relevant not only for donor-bound excitons in silicon, but potentially for other electronic states or materials as well. The extracted deformation potentials are directly relevant to the development of silicon donor-based quantum devices, particularly quantum memory and spin-photon interfaces that rely on donor-bound excitons. In these architectures, the ability to predict and tune D$^0$$\rightarrow\,$D$^0$X transition energies under magnetic fields and environmental strain is a critical requirement.
\section*{\label{sec:Acknowledgments}Acknowledgments}
The authors thank U. Gerstmann (University of Paderborn) and H. Huebl (Walther-Meißner-Institut) for valuable discussions. We thank Nikolay Abrosimov of the Leibniz-Institut für Kristallzüchtung (IKZ), Berlin, for providing the $^{28}$Si samples doped with Sb, P and As used in this study. This work was funded by the Deutsche Forschungsgemeinschaft (German Research Foundation, DFG) under Germany’s Excellence Strategy (EXC-2111, 390814868).
\appendix
\section{Additional data}
\begin{figure*}[htbp]
    \includegraphics{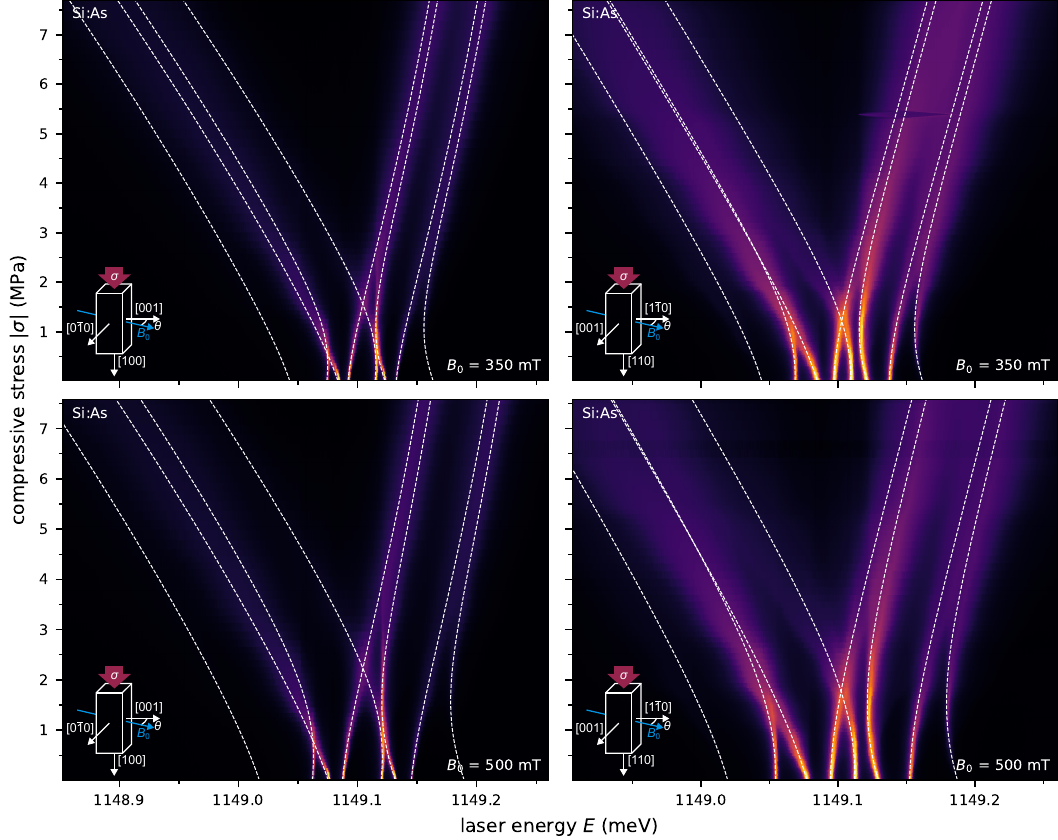}
    \caption{\label{fig:As_low_field_strain_maps}Arsenic neutral donor to donor-bound exciton transition spectra (D$^0$$\rightarrow\,$D$^0$X) in isotopically enriched $^{28}$Si under varying uniaxial stress and external magnetic fields. Each panel corresponds to a unique combination of stress orientation ([100] or [110]) and magnetic field strength (350 mT or 500 mT), as labeled in the subplots. For details regarding color code and fitted lines, refer to the caption of Fig.~\ref{fig:As_35G}.}

\end{figure*}

\begin{figure*}[htbp]
    \includegraphics{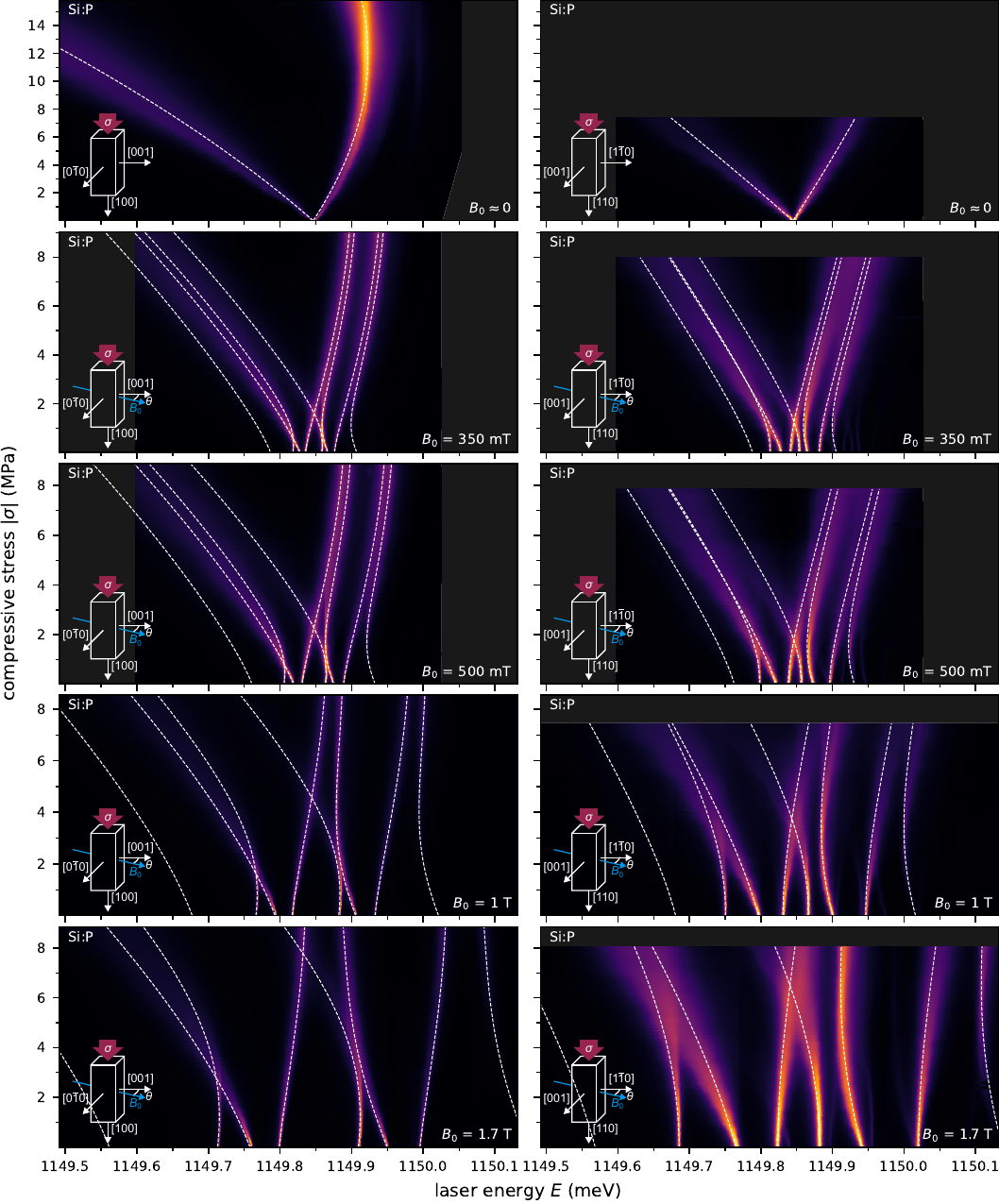}
    \caption{\label{fig:Ph_all_strain_maps} Phosphorus neutral donor to donor-bound exciton transition spectra (D$^0$$\rightarrow\,$D$^0$X) in isotopically enriched $^{28}$Si under varying uniaxial stress and external magnetic fields. Each panel corresponds to a unique combination of stress orientation ([100] or [110]) and magnetic field strength, as labeled in the subplots. For details regarding color code and fitted lines, refer to the caption of Fig.~\ref{fig:As_35G}.}
\end{figure*}

\begin{figure*}[htbp]
    \includegraphics{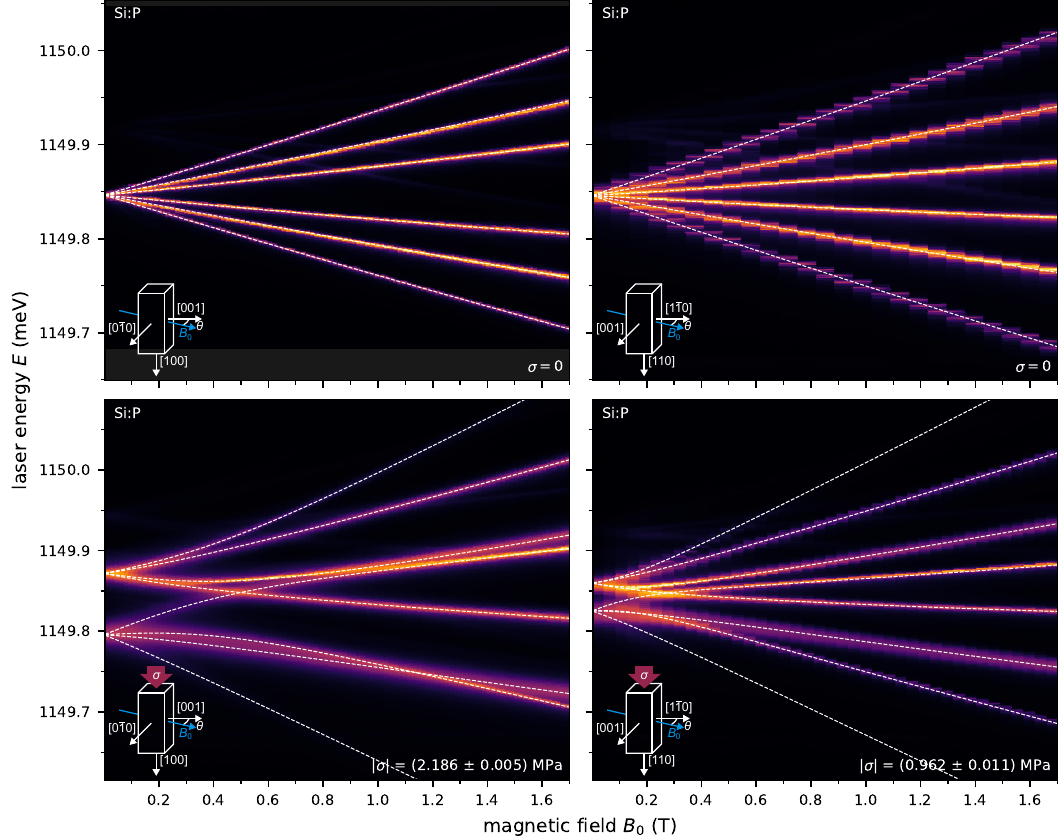}
    \caption{\label{fig:Ph_all_Zeemans}Zeeman maps of the phosphorus neutral donor to donor-bound exciton transition (D$^0$$\rightarrow\,$D$^0$X) in isotopically enriched $^{28}$Si measured under nominally zero or a small compressive stress, oriented along either [100] or [110] crystal axes, as labeled in the subplots. For details regarding color code and fitted lines, refer to the caption of Fig.~\ref{fig:As_35G}.}
\end{figure*}

\begin{figure*}[htbp]
    \includegraphics{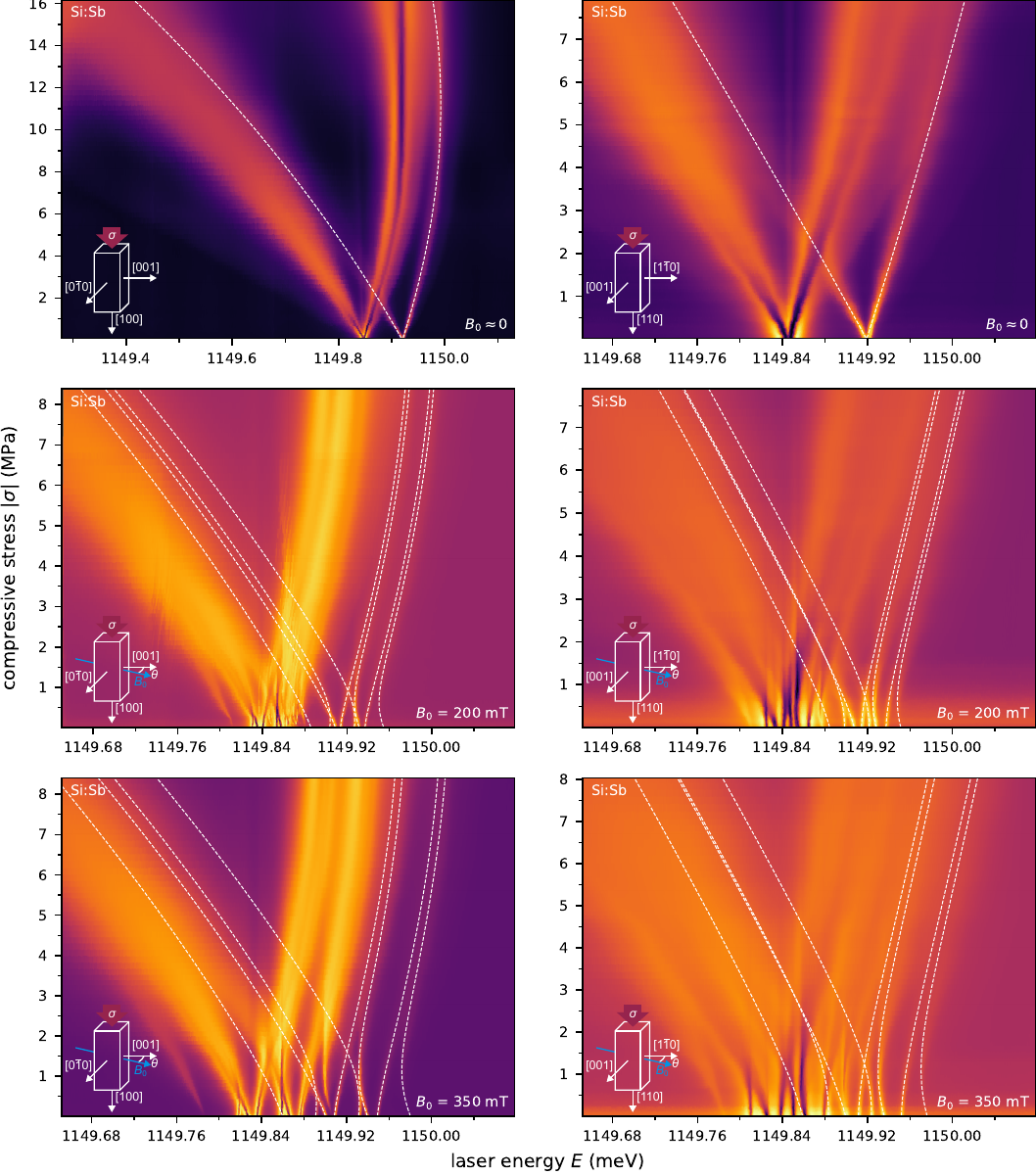}
    \caption{\label{fig:Sb_all_strain_maps} Antimony and phosphorus neutral donor to donor-bound exciton transition spectra (D$^0$$\rightarrow\,$D$^0$X) in isotopically enriched $^{28}$Si measured with higher laser power, and under varying uniaxial stress and external magnetic fields. Each panel corresponds to a unique combination of stress orientation ([100] or [110]) and magnetic field strength, as labeled in the subplots. For details regarding color code, refer to the caption of Fig.~\ref{fig:As_35G}. The much weaker antimony transitions are indicated by white dashed lines, showing a simulation of the antimony transitions based on phosphorus deformation potentials. The stronger phosphorus transitions display a higher-order artifact, visible as a narrower, inverted signal. For more details regarding the simulation of the antimony data and the measurement artifact, see Sec.~\ref{sec:Discussion}.}
\end{figure*}

\begin{figure*}[htbp]
    \includegraphics{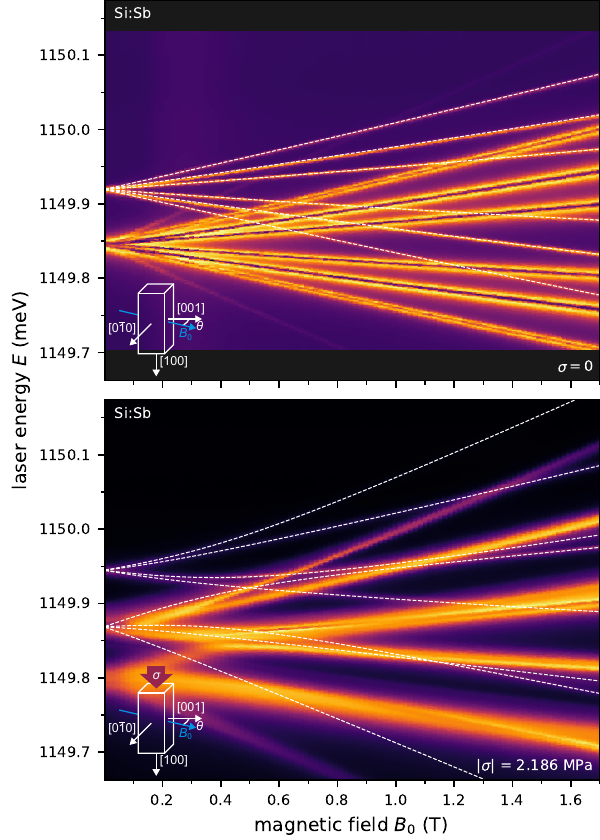}
    
    \caption{\label{fig:Sb_all_Zeemans} Zeeman maps of the phosphorus and antimony neutral donor to donor-bound exciton transitions (D$^0$$\rightarrow\,$D$^0$X) measured with higher laser power, and under [100] stress and varying magnetic fields. The two panels correspond to nominally zero or a small compressive stress, as labeled in the subplots. For details regarding color code and fitted lines, refer to the caption of Fig.~\ref{fig:As_35G}. The much weaker antimony transitions are indicated by white dashed lines, showing a simulation of the antimony transitions based on phosphorus deformation potentials. The stronger phosphorus transitions display a higher-order artifact, visible as a narrower, inverted signal. For more details regarding the simulation of the antimony data and the measurement artifact, see Sec.~\ref{sec:Discussion}.
    }
\end{figure*}

The appendix comprises five figures presenting the remaining D$^0$$\rightarrow\,$D$^0$X transition data: Figure~\ref{fig:As_low_field_strain_maps} shows As strain maps measured at lower magnetic fields, Figs.~\ref{fig:Ph_all_strain_maps} and \ref{fig:Ph_all_Zeemans} show the P strain and Zeeman maps, respectively, and Figs.~\ref{fig:Sb_all_strain_maps} and \ref{fig:Sb_all_Zeemans} show the Sb strain and Zeeman maps, measured at higher laser power to enhance the Sb signal in the presence of the strong P signal.

\clearpage
\bibliographystyle{apsrev4-2}
\bibliography{StrainedDBESpectra}

\end{document}